\documentclass[doublecol,figures]{epl2}

\newcommand{\ignore}[1]{}

\title{Charge Ordering in Half-Doped Manganites:
Weak Charge Disproportion and Leading Mechanisms}
\shorttitle{Charge Ordering in Half-Doped Manganites} 

\author{Dmitri Volja\inst{1,2} \and Wei-Guo Yin\inst{1}\thanks{E-mail: \email{wyin@bnl.gov}} \and Wei Ku\inst{1,2}\thanks{E-mail: \email{weiku@bnl.gov}}}
\shortauthor{D. Volja \etal}

\institute{
  \inst{1} Condensed Matter Physics and Materials Science Department, Brookhaven National Laboratory, Upton, New York 11973, USA\\
  \inst{2} Physics Department, State University of New York, Stony Brook,
  New York 11790, USA
}

\abstract{The apparent contradiction between the recently observed weak
charge disproportion and the traditional Mn$^{3+}$/Mn$^{4+}$ picture of
the charge-orbital orders in half-doped manganites is resolved by a
novel Wannier states analysis of the LDA$+U$ electronic structure.
Strong electron itinerancy in this charge-transfer system
significantly delocalizes the occupied low-energy ``Mn$^{3+}$''
Wannier states such that charge leaks into the ``Mn$^{4+}$''-sites.
Furthermore, the leading mechanisms of the charge order are quantified
via our first-principles derivation of the low-energy effective Hamiltonian.
The electron-electron interaction is found to play a role as important as the
electron-lattice interaction.
\ignore{A general picture of doped holes in strongly correlated
charge-transfer systems is presented and applied to the study of
charge order in half-doped manganites, using a novel Wannier states
analysis of the LDA$+U$ electronic structure. While residing
primarily in the oxygen atoms, the doped holes form additional
effective $e_g$ orbitals at the low-energy scale, leading to an
effective Mn$^{3+}$/Mn$^{4+}$ valence picture that enables weak
charge disproportion, resolving the current serious contradictions
between the recent experimental observations of charge distribution
and traditional models. Furthermore, the leading mechanisms of the
observed charge order are quantified via our first-principles derivation
of the low-energy effective Hamiltonian}%
}

\begin{document}

\maketitle

\section{Introduction}
The exploration of interplay among distinct orders lies in the heart
of condensed matter physics and materials science, as this interplay
often give rises to tunable properties of practical applications,
such as exotic states and colossal responses to external stimuli.
Manganese oxides such as La$_{1-x}$Ca$_x$MnO$_3$, which host rich
charge, orbital, spin, and lattice degrees of freedom, have thus
attracted great attention~\cite{Dagotto}. In particular, the vastly
interesting colossal magnetoresistance (CMR) effect for $x \sim
0.2-0.4$\ignore{~\cite{Jin\ignore{Schiffer}}} exemplifies rich
physics originating from proximity of competing orders. In a
slightly more doped system ($x=0.5$), all these orders coexist in an
insulating state \cite{Goodenough}, providing a unique opportunity
for a clean investigation of the strength and origin of each order.
Therefore, the study of half-doped manganites is key to a realistic
understanding of the physics of manganites in general and the CMR
effect in particular \cite{CMR:review:moreo,Akahoshi}.

The peculiar multiple orders in half-doped manganites have long been
understood in the Goodenough model \cite{Goodenough} of a Mn 3+/4+
checkerboard charge order (CO) with the occupied Mn$^{3+}$ $e_g$
orbitals zigzag ordered in the CE-type antiferromagnetic background
\cite{Goodenough}. Pertaining to CMR, it is broadly believed that a
key is the emerging of nanoscale charge-ordered insulating regions
of Goodenough type at intermediate temperatures, which could melt
rapidly in the magnetic field
\cite{CMR:review:moreo,CMR:dagotto:sen}.

Nonetheless, the simple yet profound Goodenough picture has been
vigorously challenged thanks to recent experimental observations of
nearly indiscernible charge disproportion (CD) in a number of
half-doped manganites
\cite{CMR:review:coey,CMR:lacamno3:tyson,Garcia,CMR:prcamno3:daoud-aladine,CMR:prcamno3:thomas,CMR:prcamno3:grenier,CMR:ndsrmno3:herrero-martin,CMR:prcamno3:goff}.
Such weak CD has also been observed in first-principles computations
\cite{Anisimov1,Ferrari} and charge transfer between Mn and O sites
was reported as well \cite{Ferrari}. In essence, these findings have
revived a broader discussion on the substantial mismatch of valence
and charge in most charge-transfer insulators. Indeed, extensive
experimental and theoretical effort has been made in light of the
novel Zener-polaron model \cite{CMR:prcamno3:daoud-aladine,jooss} in
which all the Mn sites become equivalent with valence being $+3.5$.
\emph{Amazingly}, most of these investigations concluded in favor of
two distinct Mn sites as predicted in the Goodenough model
\cite{CMR:prcamno3:grenier,CMR:ndsrmno3:herrero-martin,CMR:prcamno3:goff,CMR:prcamno3:trokiner,CMR:lacamno3:patterson},
calling for understanding the emergence of weak CD within the 3+/4+
valence picture.

Another closely related crucial issue is the roles of different
microscopic interactions in the observed charge-orbital orders, in
particular the relevance of electron-electron ($e$-$e$) interactions
in comparison with the well-accepted electron-lattice ($e$-$l$)
interactions. For example, $\Delta n$, the difference in the
electron occupation number between Mn$^{3+}$ and Mn$^{4+}$ states,
was shown to be small in an $e$-$e$ only picture
\cite{CMR:model:brink}; this is however insufficient to explain the
observed weak CD, since the established $e$-$l$ interactions will
cause a large $\Delta n$ \cite{CMR:LDA:popovic}. Moreover, despite
the common belief that $e$-$l$ interactions dominate the general
physics of $e_g$ electrons in the manganites
\cite{Goodenough,CMR:review:millis}, a recent theoretical study
\cite {WGYin} showed that $e$-$e$ interaction plays an essential and
leading role in ordering the $e_g$ orbitals in the parent compound.
It is thus important to quantify the leading mechanisms in the doped
case and uncover the effects of the additional charge degree of
freedom in general.

In this Letter, we present a general, simple, yet quantitative
picture of doped holes in strongly correlated charge-transfer
systems, and apply it to resolve the above contemporary fundamental
issues concerning the charge order in half-doped manganites. Based
on recently developed first-principles Wannier states (WSs) analysis
\cite{WGYin,yin,Ku} of the LDA+$U$ electronic structure in
prototypical Ca-doped manganites, the doped holes are found to
reside primarily in the oxygen atoms. They are entirely coherent in
short range \cite{HTC:model:ZRS}, forming a Wannier orbital of Mn
$e_g$ symmetry at low-energy scale. This hybrid orbital, together
with the unoccupied Mn $3d$ orbital, forms the effective ``Mn
$e_g$'' basis in the low-energy theory with conventional 3+/4+
valence picture, but simultaneously results in a weak CD owing to
the similar degree of mixing with the intrinsic Mn orbitals, thus
reconciling the current conceptual contradictions. Moreover, our
first-principles derivation of the low-energy interacting
Hamiltonian reveals a surprisingly essential role of $e$-$e$
interactions in the observed charge order, contrary to the current
lore. Our theoretical method and the resulting simple picture
provide a general framework to utilize the powerful valence picture
even with weak CD, and can be directly applied to a wide range of
doped charge-transfer insulators.

\section{Small CD vs. 3+/4+ valence Picture}

\begin{figure}[t]
\onefigure[width=0.9\columnwidth]{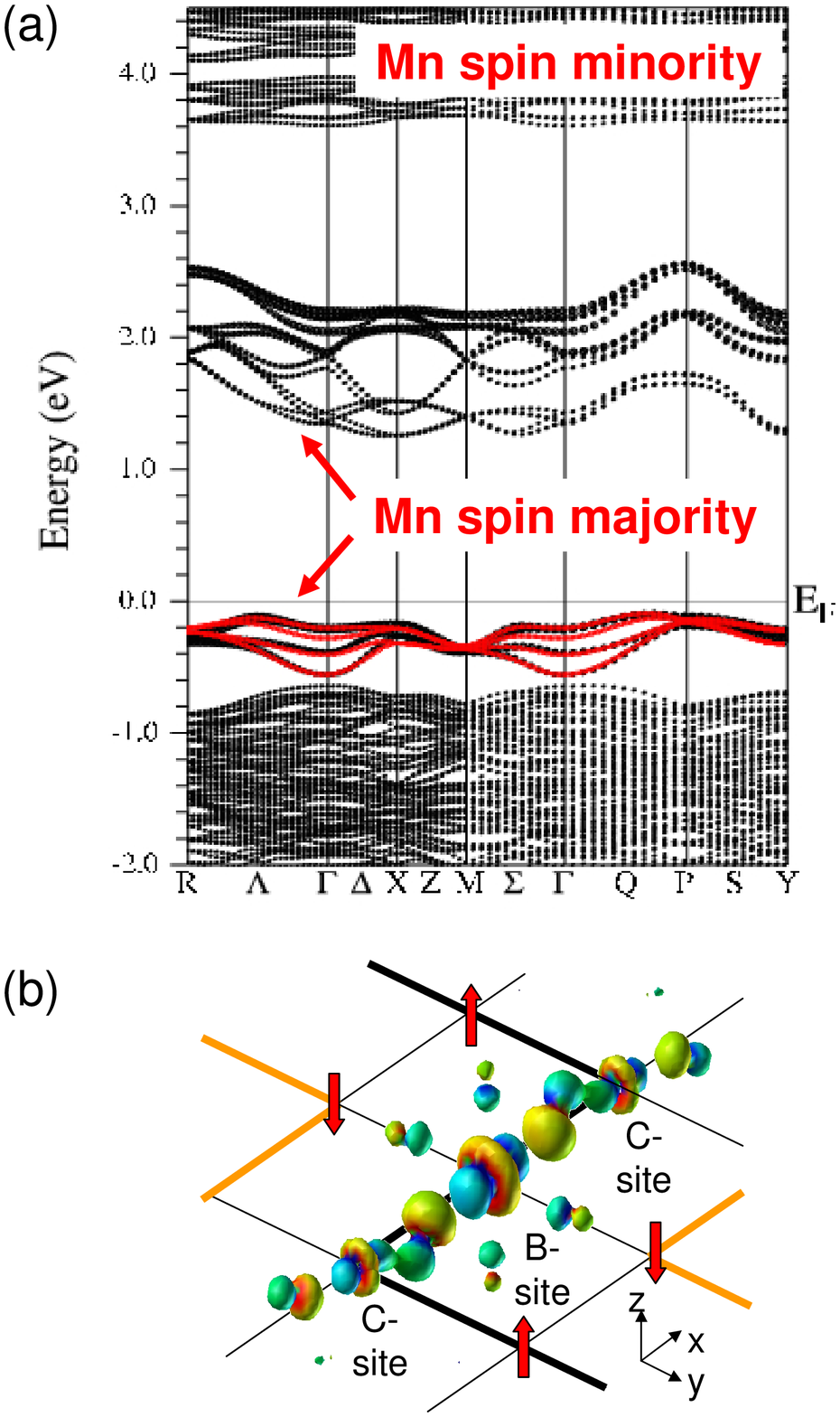}%
\onefigure[width=0.9\columnwidth]{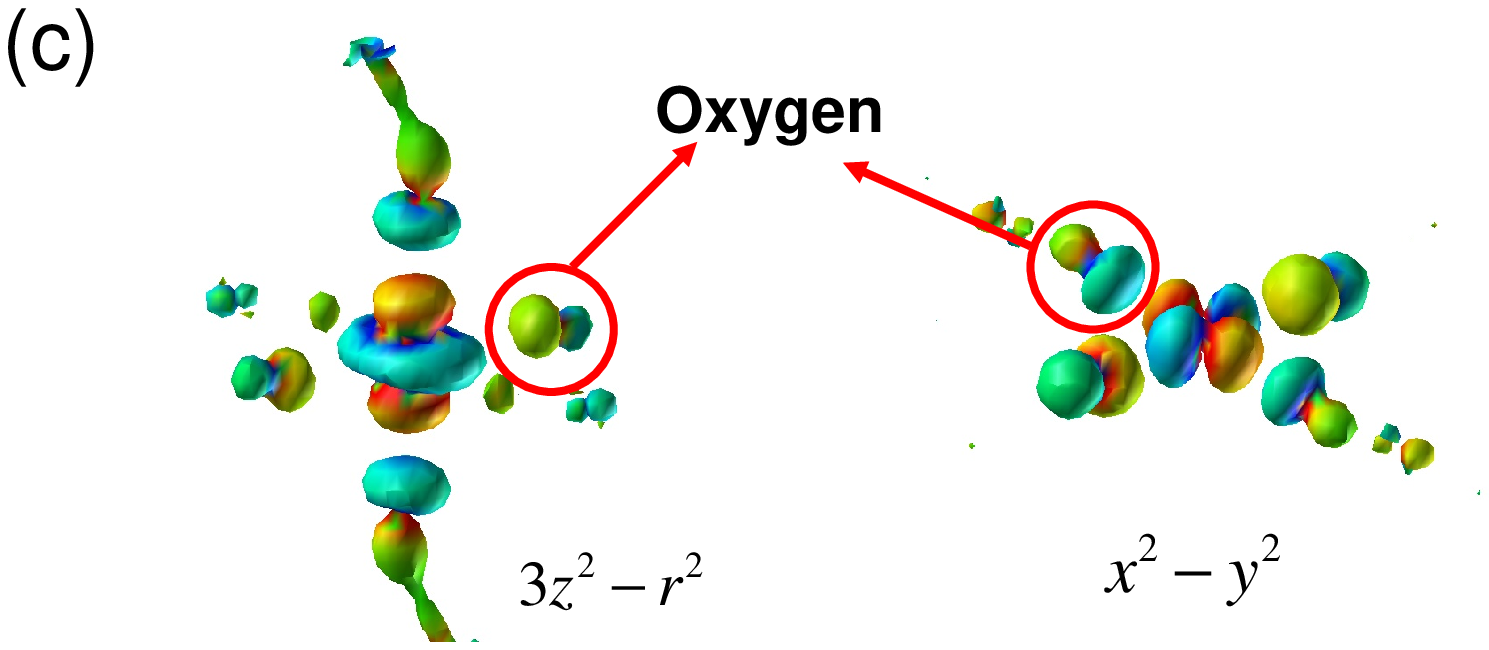}%
\caption{(Color online) (a) LDA+$U$ Band structures (dots). The (red) lines result from the
Wannier states analysis of the four occupied spin-majority Mn 3$d$-derived bands.
(b) An occupied B-site (``Mn$^{3+}$'') Wannier orbital in a spin-up (up arrow) zig-zag chain, showing
remarkable delocalization to the neighboring Mn C-sites. (c) Low-energy Mn atomic-like Wannier
states containing in their tails the integrated out O $2p$
orbitals.}
\label {ldau} %
\end{figure}

To proceed with our WSs analysis, the first-principles electronic
structure needs to reproduce all the relevant experimental
observations, including a band gap of $\sim 1.3$ eV, CE-type
magnetic and orbital orders, and weak CD, as well as two distinct Mn
sites.  We find that the criteria are met by the LDA$+U$ ($8$ eV)
\cite{Anisimov1,Blaha} band structure of the prototypical half-hoped
manganite La$_{1/2}$Ca$_{1/2}$MnO$_3$ based on the realistic crystal
structure \cite {Radaelli} supplemented with assumed alternating
La/Ca order. Hence, a proper analysis of this LDA$+U$ electronic
structure is expected to illustrate the \emph{unified} picture of
weak CD with the Mn$^{3+}$/Mn$^{4+}$ assignment, which can be easily
extended to other cases. In practice, we shall focus on the most
relevant low-energy (near the Fermi level $E_\mathrm{F}$)
bands---they are 16 ``$e_g$'' spin-majority bands (corresponding to
8 ``spin-up'' Mn atoms in our unit cell) spanning an energy window
of $~3.2$ eV, as clearly shown in Fig.~\ref{ldau}(a). For short
notation, the Mn bridge- and corner-sites in the zigzag
ferromagnetic chain are abbreviated to B- and C-sites, respectively.

The simplest yet realistic picture of the CO can be obtained by
constructing occupation-resolved WSs (ORWSs) from the four fully
occupied bands, each centered at one B-site as illustrated in
Fig.~\ref{ldau}(a). This occupied B-site $e_g$ Wannier orbital of
$3x^2-r^2$ or $3y^2-r^2$ symmetry (so formal valency is ${3+}$)
contains in its tail the integrated out O $2p$ orbitals with
considerable weight, indicative of the charge-transfer nature
\cite{Ferrari}. Moreover, this ``molecular orbital in the crystal''
extends significantly to neighboring C-sites on the same zig-zag
chain. Therefore, although by construction the C-site $e_g$ ORWSs
(not shown) are completely unoccupied (so formal valency is $4+$),
appreciable charge is still accumulated within the C-site Mn atomic
sphere owing to the large tails of the two occupied ORWSs centered
at the two neighboring B-sites. Integrating the charges within the
atomic spheres around the B- and C-site Mn atoms leads to a CD of
mere $0.14$ $e$, in agreement with experimental $0.1-0.2$ $e$
\cite{CMR:lacamno3:tyson,Garcia,CMR:prcamno3:daoud-aladine,CMR:prcamno3:thomas,CMR:prcamno3:grenier,CMR:ndsrmno3:herrero-martin}.
In this simple picture, one finds \emph{a large difference in the
occupation numbers of the ORWSs at the B- and C- sites ($\Delta
n=1$), but a small difference in real charge}. That is, the
convenient 3+/4+ picture is perfectly applicable and it allows weak
CD, as long as the itinerant nature of manganites is incorporated
via low-energy WSs rather than standard ``atomic states.''

In comparison, to make connection with the conventional atomic
picture and to formulate the spontaneous symmetry breaking with a
symmetric starting point (c.f. the next section), we construct from
the 16 low-energy bands ``atomic-like'' WSs (AWSs) of Mn
$d_{3z^2-r^2}$ and $d_{x^2-y^2}$ symmetry, as shown in
Fig.~\ref{ldau}(b). In this picture, both B- and C-site AWSs are
partially occupied with $\Delta n=0.6$. Now weak CD results from
large hybridization with O $2p$ orbitals, which significantly
decreases the charge within the Mn atomic sphere. Obviously, this
picture is less convenient for an intuitive and quantitative
understanding of weak CD compatible with the 3+/4+ picture than the
previous one, as the latter builds the information of the
Hamiltonian and the resulting reduced density matrix into the basis.
On the other hand, it indeed implies that strong charge-transfer in
the system renders it \emph{highly inappropriate to associate CD
with the difference in the occupation numbers of atomic-like
states}.

\begin{figure}[t]
\onefigure[width=0.8\columnwidth]{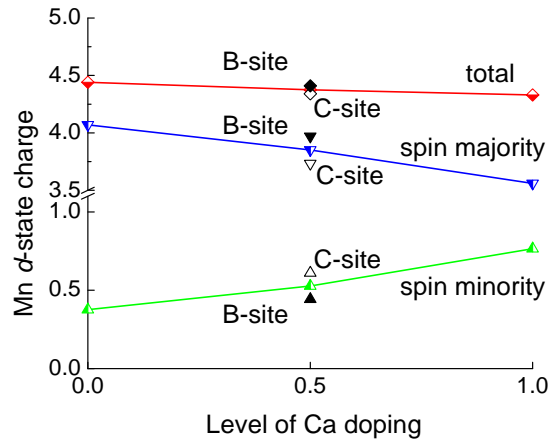}%
\caption{The calculated $d$ charge within the Mn atomic sphere for
La$_{1-x}$Ca$_{x}$MnO$_3$ ($x=0$, $1/2$, and $1$). The results are shown in terms of
(i) B-site (filled symbols), C-site (empty symbols), and site-average (half-filled symbols);
(ii) spin-majority (up triangles), spin-minority (down triangles), and total charge (diamonds).
The lines are guides to eye.}%
\label{valency}
\end{figure}

Furthermore, we find that the above conclusions are generic in
manganites by also looking into the two end limits of
La$_{1-x}$Ca$_x$MnO$_3$ ($x=0,1$). As shown in Fig.~\ref{valency},
Mn $d$ charge (within the Mn atomic sphere) is found to change only
insignificantly upon doping, in agreement with experiments
\cite{CMR:lacamno3:tyson,CMR:lasrmno4:herrero-martin}. This
indicates that doped holes reside primarily in the oxygen atoms, but
are entirely coherent in short-range and form additional effective
``Mn $e_g$'' orbitals in order to gain the most kinetic energy from
the $d$-$p$ hybridization, as shown in Fig.~\ref{ldau}(a)-(b), in
spirit similar to hole-doped cuprates \cite{HTC:model:ZRS}. This
justifies the present simplest description of the charge-orbital
orders with only the above Mn-centering $e_g$ WSs \cite{Mostovoy}.

\section{Leading Mechanisms} To identify the leading mechanisms of
the charge-orbital orders in a rigorous formalism, we proceed to
derive a realistic effective low-energy Hamiltonian,
$H^\mathrm{eff}$, following our recently developed first-principles
WS approach \cite{WGYin,yin}. As clearly shown in Fig.~\ref{ldau},
the low-energy physics concerning charge and orbital orders is
mainly the physics of one zig-zag FM chain, since electron hopping
between the antiferromagnetically arranged chains is strongly
suppressed by the double-exchange effect
\cite{CMR:model:brink,CMR:dagotto:hotta,CMR:LDA:solovyev}. Our
unbiased first-principles analysis of the 16-band one-particle
LDA$+U$ Hamiltonian in the above AWS representation reveals
\cite{CMR:model:brink,CMR:dagotto:hotta,CMR:LDA:solovyev,CMR:allen}
\begin{eqnarray}
 H^\mathrm{eff}&=&-\sum_{\langle \mathbf{ij} \rangle
\gamma\gamma^\prime}(t^{\gamma{\gamma^\prime}}_{\mathbf{ij}}d^{\dag}_{\mathbf{j}\gamma^\prime}d^{}_{\mathbf{i}\gamma}+h.c.)
-E_z\sum_{\mathbf{i}}T^z_\mathbf{i}\nonumber\\
&+&U_\mathrm{eff}\sum_{\mathbf{i}}n_{\mathbf{i}\uparrow}n_{\mathbf{i}\downarrow}
+V\sum_{\langle\mathbf{ij}\rangle}n_{\mathbf{i}}n_{\mathbf{j}}\nonumber\\
&-&g\sum_{\mathbf{i}}(\frac{1}{\sqrt{2}}\,n_\mathbf{i}Q_{1
\mathbf{i}}+T^x_\mathbf{i}Q_{2\mathbf{i}}+T^z_\mathbf{i}Q_{3\mathbf{i}})\label{heff}
\end{eqnarray}
in addition to the elastic energy $K(\{\mathbf{Q_\mathbf{i}}\})$.
Here $d^{\dag}_{\mathbf{i}\gamma}$ and $d_{\mathbf{i}\gamma}$ are
electron creation and annihilation operators at site $\mathbf{i}$
with ``pseudo-spin'' $\gamma$ defined as
$|\uparrow\rangle=|3z^2-r^2\rangle$ and
$|\downarrow\rangle=|y^2-x^2\rangle$ AWSs, corresponding to
pseudo-spin operator
$T^x_\mathbf{i}=(d^{\dag}_{\mathbf{i}\uparrow}d^{}_{\mathbf{i}\downarrow}
+
 d^{\dag}_{\mathbf{i}\downarrow}d^{}_{\mathbf{i}\uparrow})/2  $
and
$T^z_\mathbf{i}=(d^{\dag}_{\mathbf{i}\uparrow}d^{}_{\mathbf{i}\uparrow}
-d^{\dag}_{\mathbf{i}\downarrow}d^{}_{\mathbf{i}\downarrow} )/2$.
$n_\mathbf{i}=d^{\dag}_{\mathbf{i}\uparrow}d^{}_{\mathbf{i}\uparrow}
+ d^{\dag}_{\mathbf{i}\downarrow}d^{}_{\mathbf{i}\downarrow}$ is the
electron occupation number. The in-plane hoppings are basically
symmetry related: $t^{\uparrow\uparrow}_{\mathbf{ij}}=t/4$,
$t^{\downarrow\downarrow}_{\mathbf{ij}}=3t/4$,
$t^{\uparrow\downarrow}_{\mathbf{ij}}=
t^{\downarrow\uparrow}_{\mathbf{ij}}=\pm\sqrt3t/4$ where the signs
depend on hopping along the $x$ or $y$ direction. $E_z$ stands for
the oxygen octahedral-tilting induced crystal field.
$U_\mathrm{eff}$ and $V$ are effective on-site and nearest-neighbor
$e$-$e$ interactions, respectively.
$g$ is the $e$-$l$ coupling constant.
$\mathbf{Q_i}=(Q_{1\mathbf{i}},Q_{2\mathbf{i}},Q_{3\mathbf{i}})$ is
the standard octahedral-distortion vector, where $Q_{1\mathbf{i}}$
is the breathing mode (BM), and $Q_{2\mathbf{i}}$ and
$Q_{3\mathbf{i}}$ are the Jahn-Teller (JT) modes
\cite{CMR:model:brink,CMR:LDA:popovic,CMR:dagotto:hotta,CMR:LDA:solovyev,CMR:allen}.
In Eq. (1) the electron-lattice couplings have been constrained to
be invariant under the transformation of the cubic group
\cite{CMR:allen}.

The effective Hamiltonian are determined by matching its
self-consistent Hartree-Fock (HF) expression with
$H^{\mathrm{LDA}+U}$ \cite{WGYin,yin\ignore{,note:match}} owing to
the analytical structure of the LDA+$U$ approximation
\cite{VHartree}. %
\ignore{
\begin{eqnarray}
\label{mfa}
H^\mathrm{eff}_\mathrm{HF}&=&\sum_{\langle\mathbf{ij}\rangle\gamma\gamma^\prime}
(t^{\gamma{\gamma^\prime}}_{\mathbf{ij}}d^{\dag}_{\mathbf{j}\gamma^\prime}d^{}_{\mathbf{i}\gamma}+h.c.)\nonumber\\
&-&\sum_{\mathbf{i}}{(B_{1\mathbf{i}}n_\mathbf{i}+B^x_\mathbf{i}T^x_\mathbf{i}+B^z_\mathbf{i}T^z_\mathbf{i}
)}+E_0,
\end{eqnarray}
where $B^x_\mathbf{i}=2U_\mathrm{eff}\langle
T^x_\mathbf{i}\rangle+gQ^x_\mathbf{i}$,
$B^z_\mathbf{i}=2U_\mathrm{eff}\langle
T^z_\mathbf{i}\rangle+gQ^z_\mathbf{i}+E_z$ and
$B_{1\mathbf{i}}=\frac{1}{2}gQ_{1\mathbf{i}}-V\sum_\mathbf{j}{\langle}n_\mathbf{j}{\rangle}
-\frac{1}{2}U_\mathrm{eff}{\langle}n_\mathbf{i}{\rangle}$.
$E_0=U_\mathrm{eff}\sum_\mathbf{i}(\langle T^x_\mathbf{i}
\rangle^2+\langle T^z_\mathbf{i} \rangle^2-\frac{1}{4}\langle
n_\mathbf{i}\rangle^2)
-V\sum_{\langle\mathbf{ij}\rangle}{\langle}n_\mathbf{i}{\rangle}{\langle}n_{\mathbf{j}}{\rangle}+K(\{\mathbf{Q_\mathbf{i}}\})$.
}%
An excellent mapping results from $t=0.6 $ eV, $E_z=-0.08 $ eV,
$U_\mathrm{eff}=1.65 $ eV, $V=0.44 $ eV, and $g=2.35$ eV/\AA
\ignore{ \cite{note:g} \cite{note:vpotential}}. These numbers are
close to those obtained for undoped LaMnO$_3$ \cite{WGYin}
(excluding $V$, which is inert in the undoped case), and indicates
that \emph{the spin-majority $e_g$ electrons in the manganites are
still in the intermediate $e$-$e$ interaction regime with comparable
$e$-$l$ interaction}. Note that $U_\mathrm{eff}$ should be
understood as an effective repulsion of corresponding Mn-centered WS
playing the role of ``d'' states, rather than the ``bare'' $d$-$d$
interaction $U=8$ eV \cite{WGYin}. Furthermore, the rigidity of the
low-energy parameters upon significant ($x=0.5$) doping verifies the
validity of using a single set of parameters for a wide range of
doping levels, a common practice that is not \textsl{a priori}
justified for low-energy effective Hamiltonians. Clearly, the
observed optical gap energy scale of $\sim 2$ eV originates mainly
from $U_\mathrm{eff}$ instead of the JT splitting widely assumed in
existing theories \cite{CMR:dagotto:sen,CMR:review:millis}.

Now based on the AWSs, CO is measured by $\Delta n = \langle n_{\bf
i \in \mathrm{B}} \rangle - \langle n_{\bf j \in \mathrm{C}} \rangle
= 0.6$. It deviates from unity because the kinetic energy ensures
that the ground state is a hybrid of both B- and C-site AWSs, like
in the usual tight-binding modeling based on atomic orbitals.
However, since the AWSs considerably extend to neighboring oxygen
atoms, the actual CD is much smaller than $\Delta n$.

With the successful derivation of $H^\mathrm{eff}$, the microscopic
mechanisms of the charge-orbital orders emerge. First of all, note
that the kinetic term alone is able to produce the orbital ordered
insulating phase \cite{CMR:model:brink,CMR:LDA:solovyev}: Since the
intersite interorbital hoppings of the Mn $e_g$ electrons along the
$x$ and $y$ directions have opposite signs, the occupied bonding
state is gapped from the unoccupied nonbonding and antibonding
states (by $t$ and $2t$, respectively) in the enlarged unit cell. As
for orbital ordering, the Mn $d_{y^2-z^2}$ ($d_{x^2-z^2}$) orbital
on any B-site bridging two C-sites along the $x$ ($y$) direction is
\emph{irrelevant}, as the hopping integrals involving it is
vanishing. That is, only the $d_{3x^2-r^2}$ ($d_{3y^2-r^2}$) orbital
on that B-site is active and the B-sites on the zigzag FM chains
have to form an ``ordered'' pattern of alternating
$(3x^2-r^2$)/$(3y^2-r^2)$ orbitals.
However, the kinetic term alone give only $\Delta n=0$. Clearly, CO
is induced by the interactions, $U_\mathrm{eff}$, $V$, or $g$.

\begin{table}[t]
\caption{Contributions of different terms to the energy gain (in
units of meV per Mn) due to the CO formation in self-consistent
mean-field theory. BM (JT) denotes the contribution from electronic
coupling to the breathing (Jahn-Teller) mode. K denotes that from
the elastic energy.}%
\label{table:energy}%
\begin{center}
\begin{tabular}{cccccccc}
$\mathbf{Q_i}$ & Total & $U_\mathrm{eff}$  & $V$ & $t$ & BM & JT & $K$\\
\hline %
0           &  -13   & -11   & -15 & 12 & 0 & 0 & 0       \\
realistic   & -127   & -22   & -42 & 71 & -42 & -113 & 24 \\
\end{tabular}
\end{center}
\end{table}

To quantify their relative importance for CO, we calculate their
individual contributions to the total energy gain with respect to
the aforementioned $\Delta n=0$ but orbital-ordered insulating state
in the self-consistent mean-field theory. The results are listed in
Table~\ref{table:energy}. The first row obtained without lattice
distortions provides a measure of the purely electronic mechanisms
for CO. Interestingly, although the tendency is weak ($-13$ meV),
\emph{$e$-$e$ interactions all together are sufficient to induce a
CO}, consistent with the results of our first-principles
calculations. The second row is obtained for the realistic lattice
distortions, which shows a dramatic enhancement of CO by the JT
distortions ($-113$ meV), given that only half of the Mn atoms are
JT active. Together with a $-42$ meV gain from the BM distortion,
the $e$-$l$ interactions overwhelm the cost of the kinetic ($71$
meV) and elastic ($24$ meV) energy by $-63$ meV, further stabilizing
the observed CO. Nevertheless, the $-64$ meV energy gain from the
overall $e$-$e$ interactions accounts for \emph{half of the total
energy gain}, illustrating clearly their importance to the
realization of the resulting $\Delta n=0.6$. Indeed, a further
analysis reveals that $e$-$l$ couplings alone ($U_\mathrm{eff}=V=0$)
produce $\Delta n \sim 0.3$, only half of the realistic $\Delta n$,
manifesting the necessity of including $e$-$e$ interactions.

Further insights can be obtained by considering the individual
microscopic roles of these interactions acting to the kinetic-only
starting point. First, infinitesimal $U_\mathrm{eff}$ or $g$ can
induce CO, as a result of exploiting the fact that the B- (C-) site
has one (two) active $e_g$ orbital: (i) $U_\textrm{eff}$ has no
effect on the B-sites; therefore, $U_\textrm{eff}$ pushes the $e_g$
electrons to the B-sites, in order to lower the Coulomb energy on
the C-sites \cite{CMR:model:brink}. This is opposite to its normal
behavior of favoring charge homogeneity in systems such as straight
FM chains (realized in the C-type antiferromagnet). (ii) It is
favorable to cooperatively induce the $(3x^2-r^2$)/$(3y^2-r^2)$-type
JT distortions on the B-sites and the BM distortions on the C-sites
in order to minimize elastic energy \cite{Goodenough}. These lattice
distortions lower the relative potential energy of the active
orbitals in the B-sites, also driving the $e_g$ electrons there.
Hence, $U_\textrm{eff}$ and $e$-$l$ interactions work cooperatively
in the CO formation.

Unexpectedly, we find that $V$ alone must be larger than $V_c \simeq
0.8$ eV to induce CO. This is surprising in comparison to the
well-known $V_c=0$ for straight FM chains. The existence of $V_c$ is
in fact a general phenomenon in a ``pre-gapped'' system. Generally
speaking, in a system with a charge gap, $\Delta_0$, before CO takes
place (e.g. the zigzag chain discussed here), forming CO always
costs non-negligible kinetic energy due to the mixing of states
across the gap. As a consequence, unlike the first order energy gain
from $U_\mathrm{eff}$ and $g$, the second order energy gain from $V$
is insufficient to overcome this cost until $V$ is large enough (of
order $\Delta_0$). In this specific case, $V=0.44$ eV is
insufficient to induce CO by itself, but it does contribute
significantly to the total energy gain once CO is triggered by
$U_\mathrm{eff}$ or $g$, as discussed above.

It is worth mentioning that the contribution from the $E_zT_z$ term
is neglected from Table 1 because of the small coefficient of $E_z
\simeq 8$ meV, consistent with the previous study for undoped
La$_3$MnO$_3$ \cite{WGYin}. In perovskites, the tilting of the
oxygen octahedra could yield the Jahn-Teller-like distortion of
GdFeO$_3$ type, which can be mathematically described by the
$E_zT_z$ or $E_xT_x$ term. However, in the \emph{perovskite}
manganites these terms are shown here and in Ref. \cite{WGYin} to be
negligibly small. In addition, in half-doped manganites, the pattern
of the orbital order is predominantly pined by the zig-zag pathway
of the itinerant electrons and thus the effect of tilting is less
relevant. On the other hand, the $E_zT_z$ is very effective to
explain the zig-zag orbital ordering of $(x^2-z^2)$/$(y^2-z^2)$
pattern in half-doped \emph{layered} manganites, such as
La$_{0.5}$Sr$_{1.5}$MnO$_4$ Ref. \cite{huang}, where pseudo-spin-up
(the $3z^2-r^2$ orbital) is favored by a much stronger $E_z$ due to
the elongation of the oxygen octahedral along the c-axis. Described
via the pseudo-spin angel, $\theta=\arctan (T_x/T_z)$~\cite{WGYin},
$E_z$ significantly reduces $\theta$ from $\pm 120^\circ$ [i.e.,
$(3x^2-r^2)$/$(3y^2-r^2)$] to $\pm 30^\circ$ [i.e.,
$(x^2-z^2)$/($y^2-z^2)$.

The present results would impose stringent constraints on the
general understanding of the manganites. For example, Zener polarons
were shown to coexist with the CE phase within a purely electronic
modeling of near half-doped manganites \cite{Efremov}. However, to
predict the realistic phase diagram of the manganites, one must take
into account the lattice degree of freedom. In fact, when proposing
the CE phase, Goodenough considered its advantage of minimizing the
strain energy cost. Note that the previous HF theory indicated a
decrease of the total energy by $0.5$ eV per unit cell with
Zener-polaron-like displacement \cite{Ferrari}. The present LDA$+U$
calculations reveal an increase of the total energy by 1.07 eV per
unit cell with the same Zener-polaron-like displacement. This
discrepancy is quite understandable from the characteristics of the
LDA functional, which favors covalent bond, while the HF
approximation tends to over localize the orbital and disfavor
bonding.  To resolve the competition between the CE phase and the
Zener polaron phase, real structural optimization is necessary and
will be presented elsewhere. As another example, in the absence of
$e$-$l$ interactions the holes were predicted to localize in the
B-site-type region (i.e., the straight segment portion of the zigzag
FM chain) in the C$_x$E$_{1-x}$ phase of doped E-type manganites
\cite{CMR:dagotto:hotta_03}. However, since the B-sites are susceptible
to the JT distortion, they are more likely to favor electron
localization instead; future experimental verification is desirable.

\section{Summary} A general first-principles Wannier function based
method and the resulting valence picture of doped holes in strongly
correlated charge-transfer systems are presented. Application to the
charge order in half-doped manganites reconciles the current
fundamental contradictions between the traditional 3+/4+ valence
picture and the recently observed small charge disproportion. In
essence, while the doped holes primarily resides in the oxygen
atoms, the local orbital are entirely coherent following the
symmetry of Mn $e_g$ orbital, giving rise to an effective valence
picture with weak CD. Furthermore, our first-principles derivation
of realistic low-energy Hamiltonian reveals a surprisingly important
role of electron-electron interactions in ordering charges, contrary
to current lore. Our theoretical method and the resulting flexible
valence picture can be applied to a wide range of doped
charge-transfer insulators for realistic investigations and
interpretations of the rich properties of the doped holes.

\acknowledgments We thank E. Dagotto for stimulating discussions and
V. Ferrari and P. B. Littlewood for clarifying their Hartree-Fock
results \cite{Ferrari}. The work was supported by U.S. Department of
Energy under Contract No. DE-AC02-98CH10886 and DOE-CMSN.


\begin{thebibliography}{0}
\bibitem{VHartree} Unlike the local interactions, the intersite interactions
in LDA+$U$ is treated via LDA functional, which has dominant Hartree
contribution that we used for this mapping.
\end{thebibliography}
\end{document}